# Diffraction by multipoles in a $5d^2$ rhenium double perovskite

S. W. Lovesey[1,2], D. D. Khalyavin[1], G. van der Laan[2] and G. J. Nilsen[1]

[1]ISIS Facility, STFC, Didcot, Oxfordshire OX11 0QX, UK

[2]Diamond Light Source Ltd, Didcot, Oxfordshire OX11 0DE, UK

**Abstract** A recent polarized neutron diffraction experiment on the $5d^2$ rhenium double perovskite $Ba_2YReO_6$ held at a low temperature uncovered weak magnetic diffraction peaks. Data analysis inferred a significantly reduced Re dipole moment, and long-range order compatible with an antiferromagnet, non-collinear motif. To interpret the experimental findings, we present a model wavefunction for Re ions derived from the crystal field potential, Coulomb interaction, and spin-orbit coupling that fully respects the symmetry of the low-temperature ordered state. It is used to calculate in analytic form all multipole moments visible in neutron and resonance enhanced x-ray diffraction. A minimal model consistent with available neutron diffraction data predicts significant multipolar moments up to the hexadecapole, and, in particular, a dominant charge-like quadrupole moment. Calculated diffraction patterns embrace single crystal x-ray diffraction at the Re L-edge, and renewed neutron diffraction, to probe the presumed underlying multipolar order.

## I. INTRODUCTION

Scattering techniques yield a wealth of knowledge about electronic properties of materials at an atomic level of detail. To begin with, a Bragg diffraction pattern produced by Thomson scattering of x-rays is analysed in terms of spheres of electronic charge arranged on a structure defined by a space group. Space-group forbidden Bragg spots attributed to quadrupole moments produced by departures from spherical distributions of charge are weak, by their very nature. Their intensities can be enhanced by an atomic resonance in what is often called Templeton-Templeton (T & T) scattering. Classic examples are sodium bromate and sodium chlorate, which possess the same chirality yet opposite senses of optical rotation [1]. Beyond, Bragg diffraction by a non-magnetic hexadecapole observed in ichor-like haematite ($\alpha$-$Fe_2O_3$) [2]. Magnetic octupoles occur in haematite and vanadium sesquioxide ($V_2O_3$), for example, and are fully understood [3, 4]. Moreover, parity-odd magnetic (Dirac) multipoles have been observed in both materials by x-ray diffraction [5, 6]. In addition to the examples of multipoles in the two 3d-transition metal materials are multipoles observed in diffraction by rare earth and actinide (f-electron) compounds, e.g., neptunium dioxide $NpO_2$ [7] and $URu_2Si_2$ [8], and like work is reviewed by Suzuki *et al*. [9]. Neutron diffraction came late to the party with respect to gathering information on higher-order magnetic multipoles, although it is the technique of choice for determining motifs of conventional (axial) magnetic dipoles, beginning with a demonstration by Shull and Smart in 1949 of antiferromagnetic order in NaCl-type MnO below 122 K [10]. Two decades later, Moon, Riste and Koehler demonstrated advantages of exploiting neutron polarization analysis [11]. Specifically, the technique yields superior statistics on weak magnetic Bragg spots. Recent examples include the detection of long-range magnetic orders in the pseudo-gap phases of YBCO and Hg1201, and field-induced magnetization in $Sr_2IrO_4$ [12, 13]. Subsequent analyses of the patterns revealed contributions to diffraction by Dirac multipoles in the ceramic superconductors, and $5d^5$ quadrupoles and octupoles in the iridate [14, 15]. Extra knowledge, including moment directions, derived from polarization analysis refines models and, thereby, makes predictions and functional designs ever more reliable [16].

Work reported here is motivated by the observation of a magnetic powder Bragg diffraction pattern for the low-temperature (1.8 K) modification of the rhenate $Ba_2YReO_6$, which crystalizes in the elpasolite structure [17]. Albeit composed of weak Bragg features, the pattern observed by magnetic polarized neutron diffraction is beyond reasonable doubt. The result is a cautionary note on a subtraction of Bragg diffraction patterns obtained for high and low temperature modifications of a sample to estimate the magnetic signal. Its use in earlier studies of $5d^2$ double perovskites returned null results for magnetic Bragg spots [18-21]. The inferred magnetic crystal-class for $Ba_2YReO_6$ is mmm1′ (No. 8.2.25) that contains all inversions ($\bar{1}$, 1′, $\bar{1}$′), and any kind of magnetoelectric (ME) effect is prohibited. Magnetic dipoles, depicted in Fig. 1, possess an antiferromagnet, non-collinear motif represented by space group $P_Cccn$ (No. 56.375, BNS setting [22]) involving two arms of the star {(1, 0, 0) and (0, 0, 1)}.

A wave function for the ground state of rhenium $5d^2$ compatible with symmetry uses five independent coefficients, and we report corresponding multipoles from dipole to hexadecapole. We propose a minimal model with just two coefficients; one measures the orientation of magnetic dipoles, and the second is a mixing angle for non-magnetic and magnetic crystal-field states. The model can be tested by confronting it with more knowledge on the primary magnetic order parameter, and, critically, missing knowledge about charge-like quadrupoles that are likely a secondary order parameter. Indeed, symmetry-allowed magnetoelastic coupling creates T & T scattering. The diagonal component of the quadrupole is a function of the mixing angle alone, and a non-zero hexadecapole contradicts our minimal model. Resonance enhanced single-crystal Bragg diffraction of x-rays can test all aforementioned features of charge-like multipoles, for which we report relevant scattering amplitudes. The diffraction technique can also provide the orientation of magnetic dipoles, as in a previous example using an iridate [23]. Concerning neutron Bragg diffraction, even rank multipoles that result from dyadic correlations of the anapole and position operators are absent in our calculations, because crystal-field states for the atomic configuration $d^2$ in an octahedral environment belong to a J-manifold. This is unlike an iridium ion in $Sr_2IrO_4$, for which there is more than one J-manifold in crystal-field states and even rank multipoles are permitted [15].

## II. MATERIAL PROPERTIES

Two broad anomalies at temperatures ≈ 50 K and ≈ 25 K are evident in the specific heat of $Ba_2YReO_6$ as is the absence of a sharp lambda peak, which would signal long-range magnetic order [18]. The parent double perovskite crystal structure is composed of rock-salt ordered, corner-shared octahedra. Rhenium ions are in centrosymmetric sites with symmetry $m\bar{3}m$, namely, sites (4a) (0, 0, 0) in $Fm\bar{3}m$ (No. 225), and Re charge-like hexadecapoles are permitted. Lattice constant ≈ 8.3628 Å [18, 20]. Miller indices for the parent structure satisfy F-centring with $H_o + L_o$, $K_o + L_o$, $H_o + K_o$ simultaneously even. For the most part, we use the magnetic space group $P_Cccn$ in which Re ions occupy centrosymmetric sites (4e) with symmetry . . 2'/m' that forbids a dipole moment parallel to the crystal b-axis. ME and piezomagnetic effects are absent, and Dirac multipoles are forbidden. Local rhenium coordinates are denoted ($\xi$, $\eta$, $\zeta$) with $\xi = [-1, 0, 0]$, $\eta = [0, 0, 1]$, $\zeta = [0, 1, 0]$ and integer Miller indices $h = -H_o$, $k = L_o$ and $l = K_o$.

The high-spin state of the $Re^{5+}$ ($5d^2$) is $^3F$. Spin-orbit coupling is proportional to $Z^4$, where Z is the atomic number, and atomic states of the rhenium ion are assumed to be those of the total angular momentum, J. The value of the spin-orbit coupling is likely to be lower than the free-ion value, because of bonding effects known to reduce the observed magnetic moment in antiferromagnetic structures [24].

Crystal-field states for the atomic configuration $5d^2$ in an octahedral environment are well-established [25-28]. The crystal field is diagonalized before the Coulomb interaction, which is assumed large compared to the spin-orbit coupling. In octahedral symmetry, the ground state comprises three d-electron states labelled $t_2$, which take part in π-bonding with ligand ions. Irrespective of the balance between the strengths of spin coupling and crystal field, the ground state of two electrons possesses a total spin S = 1 and three-fold orbital multiplicity. Adding spin-orbit coupling leads to a ground state denoted $^3P_2$ in Ref. [28], with a total angular momentum J = 2 formed from S = 1 and a fictitious angular momentum = 1 representing the orbital triplet [25]. The ground state is composed of a doublet ($\Gamma_3$ or E level) with magnetic projections M = 0, ±2, and a triplet ($\Gamma_5$ or $T_2$) with projections M = ±1, ±2. Specifically,

$$|\Gamma_3\rangle = |0\rangle, \ (1/\sqrt{2}) [|+2\rangle + |-2\rangle]; \ |\Gamma_5\rangle = |+1\rangle, |-1\rangle, (1/\sqrt{2}) [|+2\rangle - |-2\rangle], \qquad (1)$$

where $|M\rangle = |J = 2, M\rangle$. The double direct product $\Gamma_3 \times \Gamma_3$ of the cubic group does not contain $\Gamma_4$ ($T_1$) Thus, a dipole operator has vanishing matrix elements in the $\Gamma_3$ manifold, and the doublet is non-magnetic. On the other hand, the direct product $\Gamma_3 \times \Gamma_5$ contains $\Gamma_4$ once and non-vanishing matrix elements of a dipole between the two manifolds may exist [25].

## III. MULTIPOLES AND MINIMAL MODEL

Axial (parity-even) multipoles of integer rank K are denoted $\langle T^K_Q \rangle$, where projections Q obey $-K \leq Q \leq K$, and angular brackets $\langle ... \rangle$ specify the time-average, or expectation, value of the enclosed spherical tensor operator. The property $\langle T^K_Q \rangle^* = (-1)^Q \langle T^K_{-Q} \rangle$ yields $\langle T^K_0 \rangle$ purely real. All multipoles are time-odd for magnetic neutron scattering, of course, and the dipole $\langle T^1 \rangle$ is a linear combination of spin $\langle S \rangle$ and orbital $\langle L \rangle$ moments, to a good approximation. For x-ray Bragg diffraction enhanced by a parity-even atomic resonance the time signature of $\langle T^K_Q \rangle$ depends on K alone, with K even (odd) charge-like (magnetic). In consequence, the time signature of multipoles $\sigma_\theta = (-1)^K$ for magnetic neutron and resonance enhanced x-ray diffraction.

Symmetry $2\zeta'$ at sites occupied by rhenium ions means that multipole rank and projection obey K + Q even for all K, i.e., $\langle T^K_0 \rangle = 0$ for K odd. Saturation values of multipoles $\langle T^K_Q \rangle = \langle g | T^K_Q | g \rangle$ are derived using a ground-state wavefunction,

$$|g\rangle = \beta|0\rangle + (1/\sqrt{2})\,[\alpha|+2\rangle + \alpha^*|-2\rangle] + (1/\sqrt{2})\,[\gamma|+1\rangle + \gamma^*|-1\rangle], \qquad (2)$$

and coefficients $\{|\alpha|^2 + |\beta|^2 + |\gamma|^2\} = 1$ for normalization. The time-reversal operator changes the sign of $[\gamma|+1\rangle + \gamma^*|-1\rangle]$ and $[|+2\rangle - |-2\rangle]$ that transform as $\Gamma_5$. We find,

$$\langle T^1_\xi \rangle = \sqrt{(2/5)}\,(J\|T^1\|J)\,[\sqrt{(1/3)}\,(\alpha\,\gamma^*)' + \beta'\,\gamma'], \quad \langle T^1_\eta \rangle = -\sqrt{(2/5)}\,(J\|T^1\|J)\,[\sqrt{(1/3)}\,(\alpha\,\gamma^*)'' + \beta'\,\gamma''],$$

$$\langle T^2_0 \rangle = \sqrt{(2/35)}\,(J\|T^2\|J)\,[|\alpha|^2 - |\beta|^2 - (1/2)\,|\gamma|^2], \quad \langle T^2_{+2} \rangle = \sqrt{(1/35)}\,(J\|T^2\|J)\,[2\alpha^*\,\beta' + (\sqrt{3}/2)\,(\gamma^*)^2],$$

$$\langle T^3_{+1} \rangle = \sqrt{(1/70)}\,(J\|T^3\|J)\,[-\sqrt{3}\,\alpha^*\,\gamma + 2\,\beta'\,\gamma^*], \quad \langle T^3_{+3} \rangle = -\sqrt{(1/14)}\,(J\|T^3\|J)\,\alpha^*\,\gamma^*, \qquad (3)$$

$$\langle T^4_{+2} \rangle = (1/3)\sqrt{(1/7)}\,(J\|T^4\|J)\,[\sqrt{3}\,\alpha^*\,\beta' - (\gamma^*)^2], \quad \langle T^4_{+4} \rangle = (1/6)\,(\alpha^*)^2\,(J\|T^4\|J).$$

Reduced matrix elements $(J\|T^K\|J)$ for neutron and resonant x-ray diffraction are listed in the Appendix. Real and imaginary parts of a coefficient obey the phase convention $\alpha = \alpha' + i\alpha''$ with $\alpha'$ and $\alpha''$ purely real. Notably, $\alpha'$ and $\beta$ specify $\Gamma_3$ contributions, while $\alpha''$ and $\gamma$ specify $\Gamma_5$ contributions. Multipoles of odd rank vanish for $\gamma = 0$. Magnetic dipole moments derived from $\langle T^1_\xi \rangle$ and $\langle T^1_\eta \rangle$ are,

$$\mu_\xi = \langle (L + 2S)_\xi \rangle = (4/3)\,[\,(\alpha\,\gamma^*)' + \sqrt{3}\,\beta'\,\gamma'], \quad \mu_\eta = -(4/3)\,[(\alpha\,\gamma^*)'' + \sqrt{3}\,\beta'\,\gamma''], \qquad (4)$$

with $\mu_\zeta = 0$ from symmetry.

A minimal model is achieved with $\beta = 0$, $\alpha = \cos(\chi)$ and $\gamma = \sin(\chi)\exp(i\phi)$. Evidently, the model includes the singlet ground state of the crystal field potential $[|+2\rangle + |-2\rangle]$ mixed with

$[\gamma|+1\rangle + \gamma^*|-1\rangle]$, possibly by courtesy of a quadrupole force and opposing exchange forces [29]. Multipoles for the minimal model are labelled (a) and,

$$\mu_\xi(a) = \{(2/3)\sin(2\chi)\cos(\phi)\}, \quad \mu_\eta(a) = \{(2/3)\sin(2\chi)\sin(\phi)\}. \tag{5}$$

These expressions reveal $\phi$ as the orientation of the dipole in the $\xi$-$\eta$ plane. $P_{C}ccn$ allows two orthogonal magnetic dipole components, along a and b directions of the magnetic space-group setting depicted in Fig. 1. A combination of the two components results in a non-collinear motif of dipoles, as mentioned in Section I. Both components are permitted different from zero and combined with any amplitudes, although one component might actually dominate with the second vanishingly small be comparison. Eq. (5) defines a moment direction within the $\xi$-$\eta$ plane that, indeed, is compatible with $P_{C}ccn$.

Multipoles with odd rank vanish when the mixing angle $\chi$ is set to zero. One finds,

$$\langle T^1_\xi \rangle_a = (3/2)\sqrt{(1/30)}\,(J\|T^1\|J)\,\mu_\xi, \quad \langle T^1_\eta \rangle_a = (3/2)\sqrt{(1/30)}\,(J\|T^1\|J)\,\mu_\eta,$$

$$\langle T^3_{+1} \rangle_a = -(3/4)\sqrt{(3/70)}\,(J\|T^3\|J)\,(\mu_\xi + i\mu_\eta),$$

$$\langle T^3_{+3} \rangle_a = -(3/4)\sqrt{(1/14)}\,(J\|T^3\|J)\,(\mu_\xi - i\mu_\eta), \tag{6}$$

$\langle T^2_0 \rangle_a = \sqrt{(1/70)}\,(J\|T^2\|J)\,[3\cos^2(\chi) - 1]$, $\langle T^2_{+2} \rangle_a = (1/2)\sqrt{(3/35)}\,(J\|T^2\|J)\sin^2(\chi)\exp(-2i\phi)$.

Hexadecapoles $\langle T^4_{+2} \rangle$ and $\langle T^4_{+4} \rangle$ are proportional to $(\gamma^*)^2$ and $\cos^2(\chi)$, respectively. The result for $\langle T^2_0 \rangle_a$ gives additional meaning to $\chi$. An observed magnetic moment $\mu_o = \{(2/3)|\sin(2\chi)|\} \approx 0.29$ in units of $\mu_B$ is consistent with $\chi \approx 12.9°$, and $[3\cos^2(\chi) - 1] \approx 1.85$ [17].

A second magnetic space-group that belongs to the magnetic crystal class mmm1' is included in an analysis of the observed neutron Bragg diffraction pattern in terms of states in the Chen-Balents model [17, 29]. $C_A$mca (No. 64.480) describes a collinear antiferromagnetic structure involving one arm of the star. For this case, Re ions are in centrosymmetric sites (4a), and local coordinates $(\xi, \eta, \zeta)$ are $\boldsymbol{\xi} = [-1, 0, 0]$, $\boldsymbol{\eta} = [0, -1, 0]$, $\boldsymbol{\zeta} = [0, 0, 1]$ with integer Miller indices $h = -H_o$, $k = -K_o$ and $l = L_o$. Site symmetry mm'm' requires, $\sigma_\theta (-1)^Q = +1$, and $\langle T^K_Q \rangle = (-1)^K \langle T^K_{-Q} \rangle$. In consequence, dipoles $\langle T^1_\eta \rangle = \langle T^1_\zeta \rangle = 0$, while $\langle T^1_\xi \rangle = -\sqrt{2}\,\langle T^1_{+1} \rangle$. Multipoles use $\phi = 0$ in Eqs. (5) and (6) for the minimal model.

## IV. MAGNETOELASTIC COUPLING

Magnetic space groups under consideration belong to the six-dimensional, time-odd mX$^{5+}$ irreducible representation with the $(0,0;0,\delta_4;\delta_5,0)$ order parameter direction for the $P_{C}ccn$ and $(0,0;0,0;\delta_5,0)$ for the $C_A$mca [17]. As already mentioned, $P_{C}ccn$ involves two arms of the star $\{(1, 0, 0)$ and $(0, 0, 1)\}$. There is a trilinear free-energy invariant that couples mX$^{5+}$ with the time-even representation $X^{4+}(\rho;0;0)$ associated with the third arm (0, 1, 0), $\rho\delta_4\delta_5$. In consequence, diffraction created by time-even multipoles violating F-centring is permitted, e.g., $k + l$ odd. A specific example of this type of space-group forbidden diffraction is presented in Section V.

The $X^{4+}$ representation does not enter into mechanical decomposition of the reducible representation associated with the Re site in $Fm\bar{3}m$. This means that the magnetoelastic coupling does not couple the relevant Re-displacements with this symmetry and, therefore, only higher order Re multipoles will be responsible for the scattering at space-group forbidden reflections (T & T scattering). On the other hand, the $X^{4+}$ representation does appear in the decomposition of the mechanical reducible representation associated with oxygen positions. This result implies that the magnetoelastic coupling will move oxygen ions in such a way that normal Thomson scattering will appear in the positions forbidden by the F-centring, and oxygen displacements have the propagation vector (0, 1, 0).

## V. NEUTRON DIFFRACTION

Environments at sites (0, 0, 0) and (1/2, 0, 1/2) in $P_C$ccn are related by a dyad $2_\xi$, and $2_\xi \langle T^K_Q \rangle = (-1)^K \langle T^K_{-Q} \rangle$. The electronic structure factor for diffraction is $\Psi^K_Q = [\exp(i\boldsymbol{\kappa} \cdot \mathbf{d}) \langle T^K_Q \rangle_\mathbf{d}]$, where the Bragg wave vector $\boldsymbol{\kappa} = (h, k, l)$, and the implied sum is over four rhenium sites in a unit cell. One finds,

$$\Psi^K_Q(56.375) = [1 + \sigma_\theta (-1)^{h+k}] [\langle T^K_Q \rangle + (-1)^{h+l}(-1)^K \langle T^K_{-Q} \rangle], \quad (7)$$

with $\sigma_\theta = -1$ for magnetic neutron scattering. Bulk magnetization is zero, as expected. The selection rule $h + k$ odd from antitranslation violates F-centring, and there is no nuclear scattering. The amplitude for magnetic neutron diffraction $\langle \mathbf{Q}_\perp \rangle$ is readily obtained from standard expressions, e.g., K = 1 & 3 in Eqs. (6.2) - (6.4) in Ref. [30]. With $h + k$ odd, leading-order contributions to the intermediate magnetic scattering amplitude in $\langle \mathbf{Q}_\perp \rangle = \{\mathbf{p} \times (\langle \mathbf{Q} \rangle \times \mathbf{p})\}$ are,

$$h + l \text{ even: } \langle Q_\xi \rangle \approx 6 \langle T^1_\xi \rangle, \quad \langle Q_\eta \rangle \approx \sqrt{21}\, p_\xi\, p_\eta\, [\sqrt{15}\, \langle T^3_{+3} \rangle' + \langle T^3_{+1} \rangle'],$$

$$\langle Q_\zeta \rangle \approx -4\sqrt{21}\, p_\xi\, p_\zeta\, \langle T^3_{+1} \rangle',$$

(8)

$$h + l \text{ odd: } \langle Q_\xi \rangle \approx -\sqrt{21}\, p_\xi\, p_\eta\, [\sqrt{15}\, \langle T^3_{+3} \rangle'' - \langle T^3_{+1} \rangle''],\quad \langle Q_\eta \rangle \approx 6 \langle T^1_\eta \rangle,$$

$$\langle Q_\zeta \rangle \approx -4\sqrt{21}\, p_\eta\, p_\zeta\, \langle T^3_{+1} \rangle'',$$

where the unit vector $\mathbf{p} = (\kappa_\xi, \kappa_\eta, \kappa_\zeta)/\kappa$.

Diffraction from a powder sample has an intensity,

$$I = \sum_{K,Q} [3/(K + 1)] |\langle T^K_Q \rangle|^2, \quad (9)$$

in the absence of even rank multipoles. In the present case, $K = 1$ & $3$, and $Q = \pm 1, \pm 3$. Intensity derived from Eq. (6) for the minimal model,

$$I(a) = (1/6) \mu_o^2 \{d(\kappa) + 0.170 [\langle j_2(\kappa) \rangle + (10/3) \langle j_4(\kappa) \rangle]^2\}, \quad (10)$$

with dipole intensity $d(\kappa) = [\langle j_0(\kappa) \rangle + (76/35) \langle j_2(\kappa) \rangle]^2$ displayed in Fig. 2, together with I(a). Radial integrals $\langle j_n(\kappa) \rangle$ for $Re^{5+}$ ($^3F_2$) are taken from Ref. [31]. In the so-called dipole approximation for $d(\kappa)$ the coefficient of $\langle j_2(\kappa) \rangle$ is replaced by $(2 - g)/g = 2$, where the Landé splitting factor $g = 2/3$ [30]. For observed Bragg spots [17];

$$(1, 0, 0) \; \kappa \approx 0.7513 \; \text{Å}^{-1}, \; \langle j_0(\kappa) \rangle^2 = 0.8689 \; (0.9801), \; I(a) = 0.9802$$
$$(1, 0, 1) \; \kappa \approx 1.0625 \; \text{Å}^{-1}, \; \langle j_0(\kappa) \rangle^2 = 0.7546 \; (0.9589), \; I(a) = 0.9594$$
$$(1, 0, 2) \; \kappa \approx 1.6800 \; \text{Å}^{-1}, \; \langle j_0(\kappa) \rangle^2 = 0.4913 \; (0.8883), \; I(a) = 0.8913. \quad (11)$$

with I(a) in units of $(1/6) \mu_o^2$, and values of $d(\kappa)$ are given in brackets. There is next to no difference between $d(\kappa)$ and I(a), which includes octupoles, in the range of $\kappa$ covered in the available diffraction pattern [17]. Returning to Fig. 2, the contribution from octupoles to scattering is discernible beyond $\kappa \sim 4 \; \text{Å}^{-1}$.

## V. RESONANCE ENHANCED X-RAY DIFFRACTION

Parity-even absorption events are the only ones allowed for rhenium ions in Pcccn. Dipoles and quadrupoles contribute to x-ray Bragg diffraction enhanced by an electric dipole - electric dipole (E1-E1) event, while the octupole and hexadecapole are additions to an electric quadrupole - electric quadrupole (E2-E2) event. Absorption at Re L-edges access multipoles formed with 5d-states ($L_2$ edge $\approx 11.95$ keV, $L_3 \approx 10.53$ keV). Calculated dipole strengths $[\langle 3p|R|5d \rangle / \langle 2p|R|5d \rangle] \approx 3.3$ imply that diffraction intensity is an order of magnitude larger at $M_3 \approx 2.45$ keV [32]. Unlike neutron diffraction, our x-ray multipoles do not depend on the magnitude of the scattering vector, although the E2-E2 amplitude is proportional to the square of the photon energy. Absorption at rhenium s-state edges include the K edge $\approx 71.67$ keV and $L_1$ edge $\approx 12.52$ keV, and amplitudes have relative values $\sim \{[\langle 2s|R^2|5d \rangle_{E_{L1}}]/[\langle 1s|R^2|5d \rangle_{E_K}]\}^2 \approx 0.3$.

In the general case, x-ray multipoles $\langle t^K_Q \rangle$ are calculated using Eq. (3) and reduced matrix elements in the Appendix. Celebrated sum-rules include $[\langle t^1 \rangle_{L2} + \langle t^1 \rangle_{L3}] = - \langle \mathbf{L} \rangle/(10\sqrt{2})$ [33]. Explicit results for the minimal model used throughout this section are derived from Eq. (6). Diffraction amplitudes are calculated with the electronic structure factor (7). Amplitudes in the rotated E1-E1 channel of photon polarization, labelled $\pi'\sigma$ in standard notation used in Fig. 3 [34], accesses magnetic dipoles and charge-like quadrupoles, while dipoles are absent in the unrotated $\sigma'\sigma$ channel. For Bragg spots $\kappa = (h, 0, 0)$ and enhancement at the $L_2$ edge,

$\langle t^1_\eta(11)\rangle_{L2} = -\{[26/(225\sqrt{2})]\,\mu_\eta\}$ for $h$ odd, while for $h$ even scattering is by the real part of quadrupoles, including $\langle t^2_0(11)\rangle_{L2} = \{[8/(525\sqrt{6})][3\cos^2(\chi) - 1]\}$. Corresponding multipoles at the $L_3$ edge are obtained by simple multiplication by fractions $(19/26)$ or $-(13/4)$ for dipole and quadrupole, respectively, and the cited dipole sum-rule is readily confirmed. Different magnetic information is derived from Bragg spots indexed by $(0, k, 0)$. Specifically, the dipole $\langle t^1_\xi(11)\rangle$ for $k$ odd, while charge-like scattering is also by the real part of quadrupoles. In summary, measurements of Bragg spots indexed by $(h, 0, 0)$ and $(0, k, 0)$ have potential to deliver the orientation of dipole moments in the $\xi$-$\eta$ plane and the $\Gamma_3$-$\Gamma_5$ mixing angle.

All Bragg spots indexed by $(0, 0, l)$ arise from charge-like multipoles. Space-group forbidden T & T scattering that occurs for $l$ odd is created by purely imaginary multipoles with rank K = 2 and 4. For this case, and enhancement by an E1-E1 absorption event, the amplitude in the rotated channel of diffraction,

$$F_{\pi'\sigma}(11) = -4\sin(\theta)\cos(2\psi)\,\langle t^2_{+2}(11)\rangle''. \tag{12}$$

Here, $\psi$ is the angle of rotation of the crystal about the Bragg wave vector, and the origin of the azimuthal-angle scan places the $\xi$-axis normal to the plane of scattering. Primary and secondary x-ray beams subtend an angle $2\theta$ as in Fig. 3. Continuing with our use of the minimal model, $\langle t^2_{+2}(11)\rangle''_{L2} = -(3/350)\,[(\mu_\xi\mu_\eta)/\cos^2(\chi)]$. The amplitude for diffraction enhanced by an E2-E2 event, K or $L_1$ edge, is similar,

$$F_{\pi'\sigma}(22) = 2\sqrt{(1/7)}\cos(2\psi)\,[\sqrt{3}\sin(3\theta)\,\langle t^2_{+2}(22)\rangle'' + \sin(\theta)\{1 + \sin^2(\theta)\}\,\langle t^4_{+2}(22)\rangle'']$$

$$+ 2\sin(\theta)\cos^2(\theta)\cos(4\psi)\,\langle t^4_{+4}(22)\rangle''. \tag{13}$$

This diffraction is an analogue of observations reported by Finkelstein *et al.* for haematite [2, 35]. Intensities measure the orbital angular momentum content of the d-shell, since the s-like core state is not split by the spin-orbit interaction [36]. As for numerical values in our minimal model, we find $\langle t^2_{+2}(22)\rangle'' = -\{(9/2)\sqrt{(3/7)}\,\langle t^2_{+2}(11)\rangle''_{L2}\}$, $\langle t^4_{+2}(22)\rangle'' = -\{(55/4)\sqrt{(1/7)}\,\langle t^2_{+2}(11)\rangle''_{L2}\}$, and $\langle t^4_{+4}(22)\rangle'' = 0$. In consequence, diffraction enhanced by an E2-E2 event is a straightforward test of the model.

## VI. CONCLUSIONS

In summary, we have demonstrated that the small ordered magnetic dipole moment observed in a recent neutron diffraction experiment on $Ba_2YReO_6$ implies significant multipolar moments [17]. These arise from a combination of the spin-orbit coupled J = 2 ground state and the $P_{CCCN}$ space group symmetry of the dipole ordered low-temperature state. The contribution of of each multipole may be estimated by considering a minimal model for the ground state wavefunction that includes the mixing between the $\Gamma_3$ and $\Gamma_5$ states and the dipole moment direction as parameters inferred from experimental data. When the experimental ordered dipole moment is inserted into the corresponding model expectation value, it is found that the dominant multipolar component is the charge-like quadrupole, albeit with significant contributions up to the hexadecapole. Quadrupoles in a J-manifold do not contribute to magnetic neutron diffraction, and magnetic scattering from octupoles is peaked at large scattering wave vectors κ, illustrated in Fig. 2, which together explain why no evidence of either is observed in existing neutron data. As such, the likely underlying multipolar orders in $Ba_2YReO_6$ must be probed by renewed neutron diffraction investigations or other means; resonance-enhanced x-ray diffraction at the Re L-edge is sensitive to both dipoles and quadrupoles through E1-E1 and octupoles and hexadecapoles through E2-E2 events. These experiments must, however, await the growth of single crystals.

## APPENDIX

Herewith, reduced matrix elements used in text. First, neutron diffraction [30],

$(J||T^1||J) = (2/3) \sqrt{(10/3)} [\langle j_0(\kappa) \rangle + (76/35) \langle j_2(\kappa) \rangle]$,

$(J||T^3||J) = - (6/7) \sqrt{(2/5)} [\langle j_2(\kappa) \rangle + (10/3) \langle j_4(\kappa) \rangle]$,  (A1)

with $\kappa = \{(4\pi/\lambda) \sin(\theta)\}$, where θ is the Bragg angle displayed in Fig. 3 and λ the wave length. Radial integrals are defined such that $\langle j_0(0) \rangle = 1$, and $\langle j_n(0) \rangle = 0$ for n ≥ 2.

Turning to resonance enhanced x-ray diffraction [34],

$(J||t^1(11)||J)_{L2} = - (52/45)\sqrt{(1/15)}$, $(J||t^1(11)||J)_{L3} = - (38/45)\sqrt{(1/15)}$,

$(J||t^2(11)||J)_{L2} = (8/15) \sqrt{(1/105)}$, $(J||t^2(11)||J)_{L3} = - (26/15) \sqrt{(1/105)}$.  (A2)

Lastly, for E2-E2 reduced matrix elements (*n*s-edge, valence d-state) we use [34],

$(J||t^K(22)||J) = (1/5) \sqrt{[2(2K + 1)]} (-1)^K W^{(0K)K}$,

$W^{(0K)K} = (2J + 1) \begin{Bmatrix} S & S & 0 \\ L & L & K \\ J & J & K \end{Bmatrix} W^{(0K)}$ with $W^{(0K)} = \sqrt{[(1/2) (2S + 1)]} V(K)$,

where the Racah unit tensor V(K) is tabulated in Ref. [35]. For the atomic configuration $^3F_2$,

$(J||t^1(22)||J) = - (4/5) \sqrt{(1/3)}$, $(J||t^2(22)||J) = - (12/35) \sqrt{(1/5)}$,

$(J||t^3(22)||J) = (3/5) \sqrt{(1/7)}$, $(J||t^4(22)||J) = (11/35)$.  (A3)

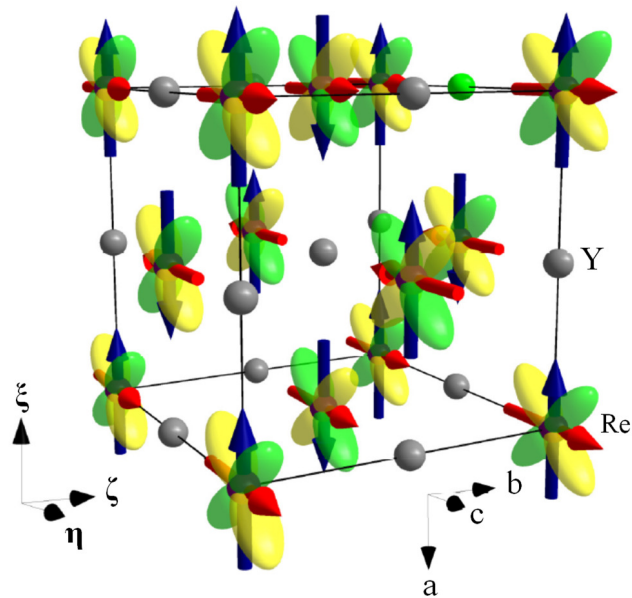

Fig. 1. Rhenium magnetic dipoles in the ξ-η plane (components red & blue arrows) and charge-like quadrupoles ∝ ξη in $Ba_2YReO_6$ using space group $P_cccn$ (No. 56.375). Grey spheres represent Y ions.

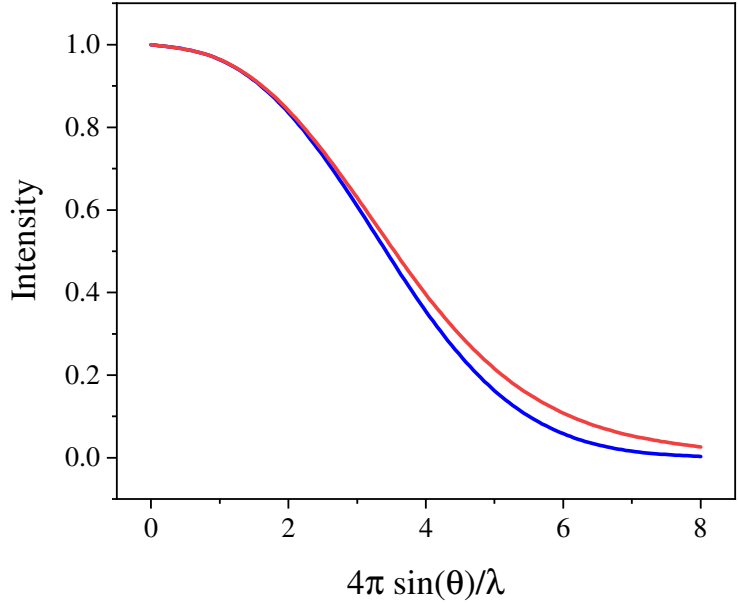

Fig. 2. Powder intensity Eq. (10) for the minimal model is displayed as a function of $\kappa = \{4\pi \sin(\theta)/\lambda\}$, where $\theta$ is the Bragg angle in Fig. 3, in the range 0 - 8 Å$^{-1}$ using $[(1/6)\,\mu_o^2]$ as a unit of intensity. Blue curve: dipole intensity $d(\kappa) = [\langle j_0(\kappa)\rangle + (76/35)\,\langle j_2(\kappa)\rangle]^2$. Red curve: dipole plus octupole contributions with $K = 3$, $Q = \pm 1, \pm 3$. Radial integrals for Re$^{5+}$ (5d$^2$) from Kobayashi et al. [30].

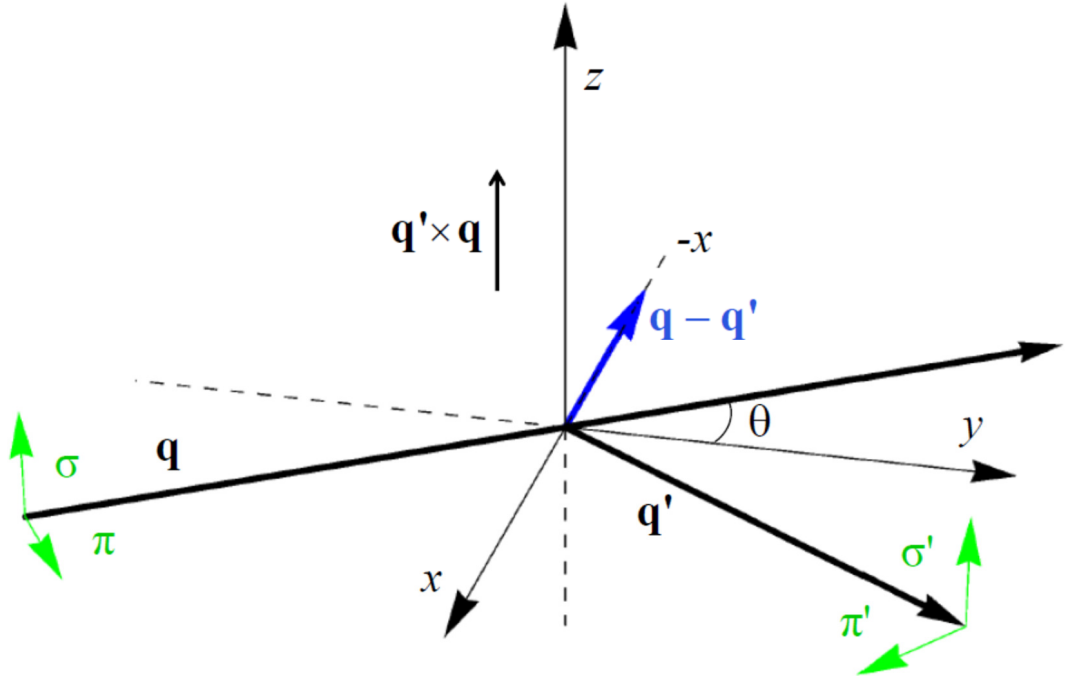

Fig. 3. X-ray diffraction. Primary ($\sigma$, $\pi$) and secondary ($\sigma'$, $\pi'$) states of polarization. Corresponding wave vectors **q** and **q'** subtend an angle $2\theta$. Local rhenium axes ($\xi$, $\eta$, $\zeta$) and depicted Cartesian co-ordinates ($x$, $y$, $z$) coincide in the nominal setting of the crystal.